# Edge effect on resistance scaling rules in graphene nanostructures

*Guangyu Xu[1,*], Carlos M. Torres Jr.[1,+], Jianshi Tang[1,+], Jingwei Bai[2,+], Emil B. Song[1], Yu Huang[2], Xiangfeng Duan[3], Yuegang Zhang[4,*] and Kang L. Wang[1]*

[1]Department of Electrical Engineering, [2]Department of Material Science and Engineering, [3]Department of Chemistry and Biochemistry, University of California at Los Angeles, Los Angeles, California 90095, USA

[4]Molecular Foundry, Lawrence Berkeley National Laboratory, 1 Cyclotron Road, Berkeley, California 94720, USA

[+]These authors contributed equally to this work

[*]email: guangyu@ee.ucla.edu (GX); yzhang5@lbl.gov (YZ)

**RECEIVED DATE ()**

**Abstract** We report an experimental investigation of the edge effect on the room-temperature transport in graphene nanoribbon and graphene sheet (both single-layer and bilayer). By measuring the resistance scaling behaviors at both low and high carrier densities, we show that the transport of single-layer nanoribbons lies in a strong localization regime, which can be attributed to an edge effect. We find that this edge effect can be weakened by enlarging the width, decreasing the carrier densities or adding an extra layer. From graphene nanoribbon to graphene sheet, the data show a dimensional crossover of the transport regimes possibly due to the drastic change of the edge effect.

Advancement of nanoscale materials has attracted considerable interest in understanding their low dimensional transport[1, 2]. Reduction of the sample size can lead to rich size-driven effects on the transport through the change of the system dimensionality[3, 4]. Graphene is an extraordinary 2D material with great potential[5, 6]. As the width of graphene narrows down to a nanometer size, graphene nanoribbon (GNR) would exhibit quasi-1D transport with the presence of an energy gap, which benefits switching on/off the devices[7, 8]. Unlike carbon nanotube with a perfectly enclosed structure[1], as-made GNR usually has unavoidable edge disorders[9]. The question of how the edge disorder affects GNR transport is both of fundamental interest and practical concern for device implementations.

Many efforts have been made to explore the edge effect on GNR transport, while no consensus has been reached[10-16]. Until now, most experiments focused on the low-temperature transport in single-layer GNR (SLR) at low carrier densities. For example, Han et al. has reported the size-scaling of the SLR transport, suggesting the origin of transport gap from a combination of the edge effect and the Coulomb charging effect[10]. However, their fabrication method leaves chemical coverage/residues on top of the samples, which makes it difficult to probe the intrinsic SLR properties. Comparatively, another work on SLRs fabricated by a metal-mask etching method attributes the size-scaling of the transport gap to the effect of charge impurities rather than that of the edge[11]. Given the sensitivity of the SLR to the weight of multiple types of scatterings, it may not be surprising to see this inconsistency in the role of edge effect on transport properties, which can be quite different in samples prepared using different methods [11, 14, 15].

Length-dependence of the resistance (i.e. resistance scaling, R-L relation) has been broadly used to probe the transport properties of electronic materials[17-20]. For instance, this method has been employed to explore the electron-phonon scattering in carbon nanotubes (CNT) by comparing the R-L curves under low-bias and high-bias conditions[17]. Furthermore, several works have used the R-L relations of CNTs to identify their transport regimes, such as the exponential R-L relation and linear R-L relation at localization and diffusion regimes, respectively[18, 19]. Fundamentally, the length-dependence of the resistance is a representation of the one-parameter scaling law, which has attracted much interest in understanding the phase transition in low-dimensional materials[20-22].

In this letter, we aim to understand the edge effect on the room-temperature transport of GNRs and graphene sheets (with micron-sized width) through investigating the different resistance scaling rules in these graphene nanostructures. GNRs are fabricated by a nanowire-mask etching method with good performance as reported before[23]. For practical concerns, both off- and on-state resistance data are collected to probe the room-temperature transport at both low and high carrier densities. Our data show that the SLR transport lies in a strong localization regime, which can be attributed to a strong edge effect. We find that this edge effect can be weakened by enlarging the width, decreasing the carrier densities or adding an extra layer. From GNR to graphene sheet, the data exhibit a dimensional crossover of the transport regimes possibly due to the drastic change of the edge contribution. This work pinpoints the critical role of the edge effect on the crossover of the transport regimes in graphene through the resistance scaling rules; this result may provide insight on realizing scalable graphene electronics.

Our graphene sheets (with micron-sized width) were mechanically exfoliated from natural graphite onto a thermally-grown 300nm $SiO_2$ dielectric layer on a highly-doped Si substrate which acts as the backside gate[7]. Subsequently, some graphene sheets are directly patterned to Hall-bar or multi-probe devices using electron beam lithography with Ti/Au electrodes[24]; while some graphene sheets are firstly etched to GNRs by oxygen plasma using a nanowire-mask method, and then patterned into multi-probe devices[25]. To avoid extrinsic doping effects, 1) the nanowire-mask on top of GNRs is removed by weak sonication before the device formation; 2) all graphene samples are maintained in vacuum and go through a baking process to desorb the contaminants before the resistance measurement. The number of graphene layers is identified through Raman spectroscopy before patterning into devices[24]. We thus have the samples made of single-layer and bilayer GNRs (SLR/BLR) and graphene sheets (SLG/ BLG). The sample dimensions (length (L) and width (W)) are determined by atomic force microscopy (AFM) for GNRs[23] (see Fig. 1a) and optical microscopy for graphene sheets, respectively.

DC resistance (R=V/I) of samples is measured within the low-bias regime at each gate bias ($V_g$). An ambipolar transport is typically observed (see R-$V_g$ curves in Fig. 1b) in SLR and SLG at both electron-conduction (nFET) and hole-conduction sides (pFET). To investigate the transport at both low and high carrier densities, we measured the resistances at both off- and on-states (i.e. $R_{off}$ and $R_{on}$). Here we define: 1) the off-state as $V_g$ at the Dirac point ($V_{Dirac}$) (the region at low carrier densities); 2) the on-state as $V_g$ at $|V_g - V_{Dirac}| \sim 30V$ (the region at high carrier densities), where R becomes near-saturated (<10% variation with further increase of $|V_g - V_{Dirac}|$). In the experiment, we confirmed that all devices are biased in the near-linear regime at both the on- and off- states to assure that the definition of R is valid (see Fig. 1c). The linear I-Vs for GNRs with W>25nm indicate that the possible Schottky-barrier near the GNR-metal contact may be smeared by temperature[7, 11, 12]. In the rest of this work, we present the data from the hole-conduction side only since those from the electron-conduction side show similar trends. All resistance data are scaled by 1/W for comparison among samples with different widths.

To investigate the SLR transport at low carrier densities, Figure 2a shows the length-dependence of the off resistance (i.e. $R_{off}$- L relation) in SLRs with W~45nm and W~34nm. The $R_{off}$- L curves do not follow a linear relation (R ∝ L), indicating that the SLR transport does not lie in a diffusive regime[17, 19]. Instead, $R_{off}$ exhibits an exponential increase with L, featuring a transport regime of strong localization[15, 20]. The strong localization observed at room temperature indicates that the inelastic scattering is weak in our SLRs, which has been similarly found in single-walled carbon nanotube (SWNT) before[18, 26]. It is interesting to point out that both SWNT and SLR feature 1D/quasi-1D transport with a single layer of carbon atoms. Although SLRs unavoidably have edge disorders whereas SWNTs do not (structural defects are shown to result in the strong localization in SWNT[18]), these edge disorders can be mostly involved in the elastic scattering and contribute little to the dephasing processes[20, 27] Fitting the data as R (L) ~ exp (L/$L_0$) with $L_0$ being the localization length[18], we find that the localization length ($L_0$) becomes smaller with a smaller width. This W-dependence of $L_0$ suggests the relevance of the edge effect on the carrier localization in SLR, since the carriers in narrower SLRs are more affected by the edge[15]. It has been predicted that the edge disorder can contribute to the non-uniformity of the local density-of-states, which leads to the carrier localization[15, 16, 28]. In this picture, this edge effect can be stronger as W becomes smaller, resulting a stronger carrier localization with a smaller $L_0$. We note that the R·W value for W~45nm is typically larger than that for W~34nm, which may also relate to the significant edge effect in our SLRs.

To testify the edge effect on the carrier localization, we can compare the SLR transport at different carrier densities via the resistance scaling. This is because the edge disorders mainly contribute to the short-range scatterings, whose weight can change with the carrier densities[27, 29, 30]. We thus extend the analysis to the on-state (i.e. $R_{on}$ - L relation) to describe the SLR transport at high carrier densities. Figure 2b shows that the on-state resistance ($R_{on}$) also exhibits an exponential R-L relation with a similar W-dependence of $L_0$ ($L_0$ is smaller at a narrower W), suggesting that the edge disorder plays a role in the carrier localization at high carrier densities. For both W~ 45nm and 34nm, we find that the $L_0$ values at the on-state are smaller than those at the off-state, indicating a stronger localization at higher carrier densities. At high carrier densities, it has been shown that the weight of edge-induced short-range scattering to the transport is larger whereas the effect of long-range disorder is insignificant due to the carrier screening[29, 30]. Hence, the trend of $L_0$ versus the carrier density further supports the edge effect as the main reason of the carrier localization: the edge effect is stronger at high carrier densities, leading a stronger localization with a smaller $L_0$. Conversely, the data show that long range disorders act to weaken the localization, because its effect is stronger at low carrier densities where the localization is weaker (with a larger $L_0$)[29]. The stronger carrier localization at the on-state in our SLRs supports the claim that the short-range scattering assists the carrier localization while the long-range scattering tends to delocalize it[27] (Note: the impact of other short-range disorders away from the edges (e.g. structural defects) should be small as indicated by the negligible D-peak intensity in Raman spectroscopy). The comparison of $L_0$ at the on- and off-state indicates that the carrier densities affect the weight of edge effect in SLR.

So far, we attribute the resistance scaling of our SLRs to the strong localization induced by the edge effect. However, we do not exclude the possibility that SLRs fabricated by other methods can behave differently, since the weight of multiple scatterings in the fabricated SLRs can be quite different depending on the fabrication methods[11]. For example, HSQ-based patterning methods can leave HSQ coverage/residual which dopes the SLRs[10]; top-gate structure or the coverage of the etching mask can act as an extra SLR-dielectric interface which causes scattering/screening to the carriers[10, 31]. The large variability in SLRs is similar to that of SWNTs by various growth/fabrication methods[18, 26]. In the future, the role of edge effect on resistance scaling behaviors in SLRs could be further explored by edge-engineering or changing their substrates.

We next examine the resistance scaling in BLRs. Comparing with SLR, BLR transport is not well understood yet and the experimental works are rare[23, 32]. We thus limit the discussions on the effect of the extra layer of BLR (compared to SLR) to the transport by R-L relations (see Fig. 3a and Fig. 3b). The main feature of BLR is that: in contrast to SLR, the resistances of BLR at both low and high carrier densities ($R_{off}$ and $R_{on}$) exhibit a linear increase with L (i.e. $R \propto L$), characteristic of diffusive transport instead[17]. For BLRs, it is suggested that the short-range edge disorder should play a more important role to the transport than SLR, since the effect of long-range disorders is weaker due to a stronger screening effect in its bilayer structure[30, 33]. Hence, the absence of localization (as indicated by the R-L relation) indicates that the edge effect in BLR is weaker than that in SLR. To gain some insight on this difference, we note that some edge states of BLR have been predicted to exist only in one layer[34, 35]; the carriers on the other layer may be much less affected by these edge states. Also, the carriers in the layer with these edge states can hop to the other layer assisted by interlayer coupling[30, 33]. The effect of these edge effects can thus be weakened, with the carriers being more delocalized.

To further understand the weakened edge effect in BLR, we fit the data as $\rho = dR/dL = (h/2e^2) \cdot (1/L_m)$ where $L_m$ is the mean-free-path in the diffusive transport regime[17, 36]. The result shows that $L_m$ at the on-state is larger than that at the off-state, indicating less carrier scatterings at higher carrier densities. We note that this trend of $L_m$ versus the carrier density in our BLRs cannot be fully explained by the edge-induced short-range scattering, whose effect should lead to a smaller $L_m$ at on-state as opposed to our data[37, 38]. The larger $L_m$ at on-state in our BLRs rather appears to originate from the weakened long-range scattering at higher carrier densities similar to the case in SLR[29, 37]. Moreover, we observe a weaker W-dependence of both $L_m$ and R·W values than those in SLR (see Fig. 3a and 3b), which also supports the weak edge effect on BLR transport. All these facts pinpoint that the edge effect in BLR is weakened by adding the extra layer, as indicated by apparently different resistance scaling rules from those in SLR.

One can expect that the edge effect in SLR and BLR could be further weakened as we significantly increase their widths to form SLG and BLG. To see if the change of edge effect can induce a crossover of the transport regimes, we studied the R-L relations in SLG and BLG whose widths are typically larger than1μm. Figure 4a shows that SLG exhibits a linear R-L relation at both low and high carrier densities ($R_{off}$ and $R_{on}$), featuring a diffusive transport instead of the strong localization in SLR. This dimensional crossover from quasi-1D SLR to 2D SLG reaffirms our claim that the localization in SLR is dominated by the short-range edge disorder, whose effect is much weaker in SLG. The absence of localization in SLG also indicates that the long-range disorder (as suggested being the dominate scattering in SLG[29, 30]) cannot lead to the carrier localization. Figure 4b shows that BLG also feature a diffusive transport as indicated by the linear R-L relations at both the on-and off-states. This can be explained as BLG having an even weaker edge effect (due to the large width) than BLR; hence neither can form the localization. We fit the SLG and BLG data as $\rho_{2D} = d(R·W)/dL$ according to the 2D diffusive transport[39]. The obtained $\rho_{2D}$ values are reasonable for SLG and BLG with the hole mobility ~ 1000-6000cm$^2$/(V·s) (see Fig. 4a and Fig. 4b)[40].

In summary, we present the resistance scaling in both quasi-1D and quasi-2D graphene materials (SLR, BLR, SLG, BLG, see Fig. 5), which pinpoints the critical role of the edge effect on the crossover of the transport regimes. By measuring the R-L relation at both low and high carrier densities, we find that the SLR transport lies in the strong localization regime, which can be mainly attributed to the effect of edge-disorders. Through the comparisons among the four graphene nanostructures, we find that the edge effect on the graphene transport can be weakened by enlarging the width, decreasing the carrier densities or adding an extra layer. Our results reveal the critical role of edge effect on graphene transport and thus the resistance scaling rules, which may provide insight to realize the ultimate goal of scalable graphene electronics.

**Acknowledgement** The authors thank F. Miao, X. Zhang, R. Cheng and L. Liao for helpful discussions. We especially thank M. Y. Han from P. Kim's group for theoretical discussions. We greatly appreciate technical support from M. Wang, C. Zeng, S. Aloni and T. Kuykendall. This work was in part supported by MARCO Focus Center on Functional Engineered Nano Architectonics and the U.S. Department of Energy under Contract No. DE-AC02-05CH11231.

**Figure Captions**

**Figure 1 Typical characterization and resistance measurement of graphene nanostructures at room temperature. a.** the AFM image of a typical SLR between Ti/Au metal contacts (7/80nm). The

scale bar equals 0.3μm. **b.** R vs. ($V_g$-$V_{Dirac}$) for a SLR and a SLG with the definition of the on- and off-states. SLR is typically more resistive than SLG by 1-2 orders of magnitude. **c.** Near-linear I-V curves in a SLR when the gate is biased both near and away from the Dirac point (i.e. off- and on-states, respectively).

**Figure 2 Resistance scaling ($R_{on}$/$R_{off}$ versus L) for SLR at room temperature. a.** $R_{off\_SLR}$ exponentially increases with L. The fitting shows a characteristic localization length of $L_0$~0.27μm for W~45nm and $L_0$~0.056μm for W~34nm, respectively. The variation of widths among samples is less than 5nm. **b.** $R_{on\_SLR}$ exponentially increases with L. The fitting shows $L_0$~0.14μm for W~45nm and $L_0$~0.050μm for W~34nm, respectively.

**Figure 3 Resistance scaling ($R_{on}$/$R_{off}$ versus L) for BLR at room temperature. a.** $R_{off\_BLR}$ linearly increases with L. The fitting shows a characteristic mean-free-path of $L_m$~40nm for W~42nm and $L_m$~32nm for W~53nm. The variation of widths among samples is less than 5nm. **b.** $R_{on\_BLR}$ linearly increases with L. The fitting shows $L_m$~72nm for W~42nm and $L_m$~94nm for W~53nm, respectively.

**Figure 4 Resistance scaling ($R_{on}$/$R_{off}$ versus L) for SLG and BLG at room temperature. a.** Both $R_{on\_SLG}$ and $R_{off\_SLG}$ linearly increase with L, and have a fitted resistivity of $\rho_{on\_SLG}$ ~5.0KΩ and $\rho_{off\_SLG}$ ~7.4KΩ, respectively. **b.** Both $R_{on\_BLG}$ and $R_{off\_BLG}$ linearly increase with L, and have a fitted resistivity of $\rho_{on\_BLG}$ ~3.4KΩ and $\rho_{off\_BLG}$ ~5.7KΩ, respectively.

**Figure 5 Schematics for the crossover of transport regimes in graphene nanostructures.** The edge effect in SLR can be weakened by either adding an extra layer to form BLR or increasing the width to form SLG; both cause the transition of transport regimes from localization to diffusion. Note that the carrier densities can also affect the edge effect in SLR, which can be tuned by gate biases.

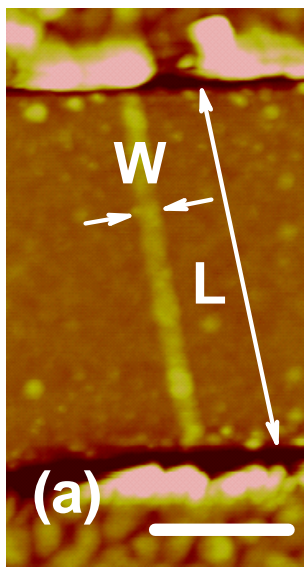

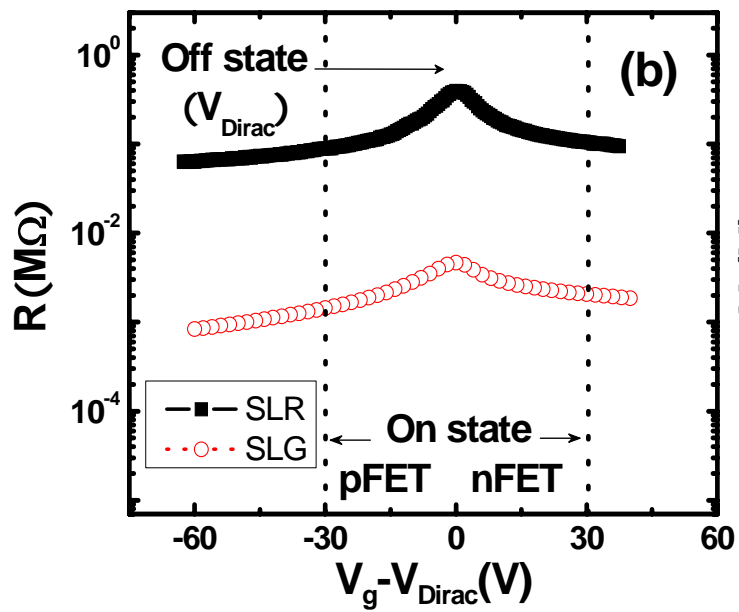

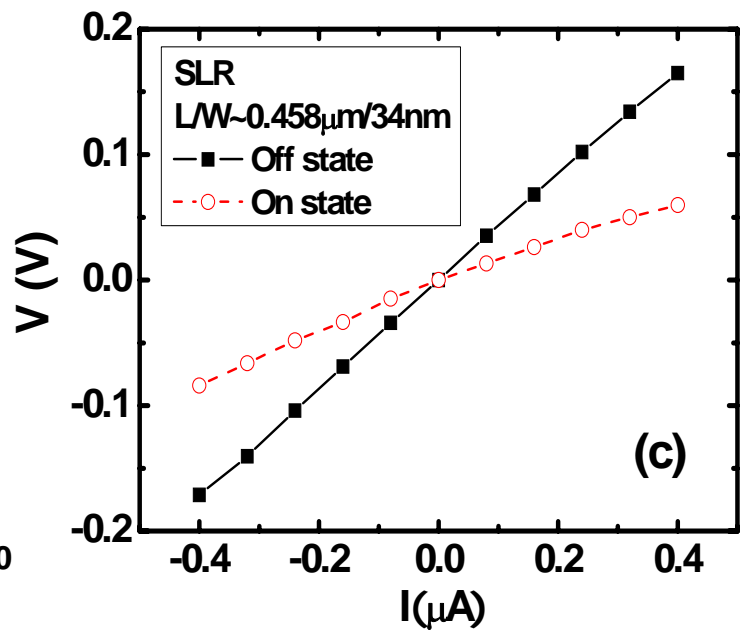

Xu Figure 1

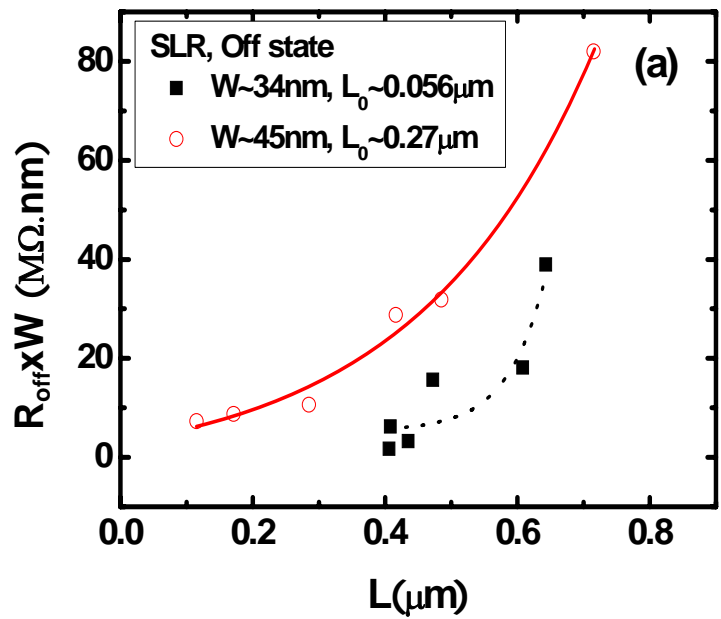
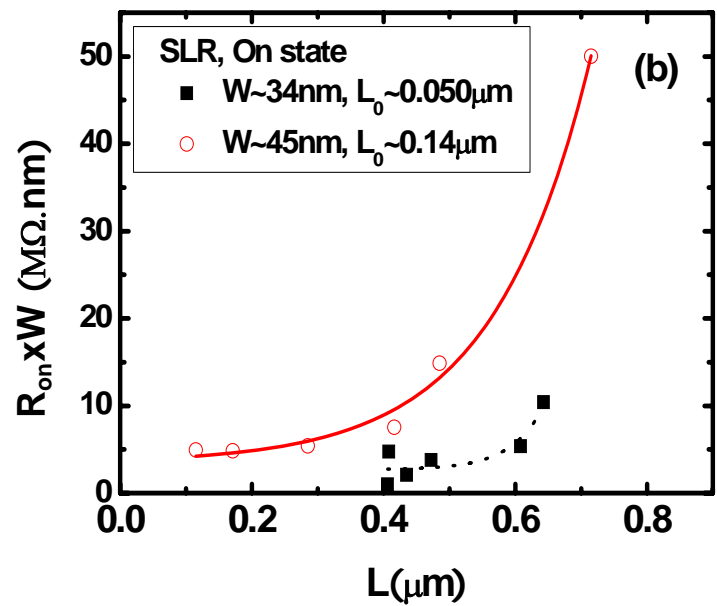

Xu Figure 2

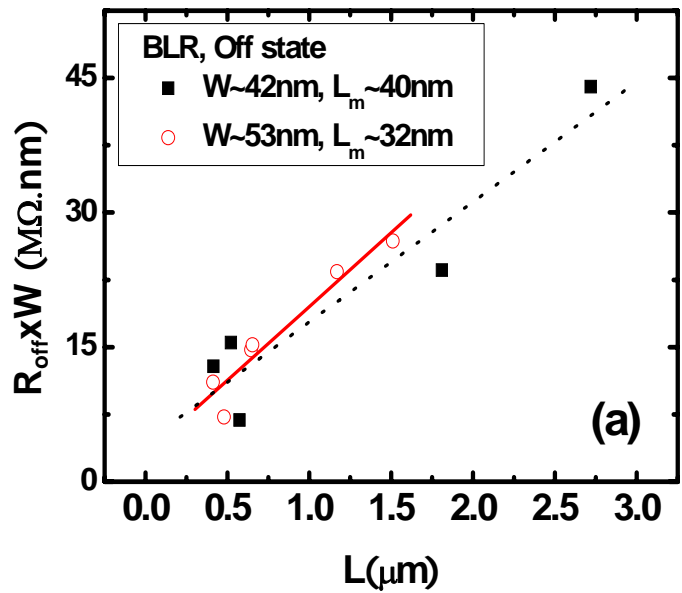
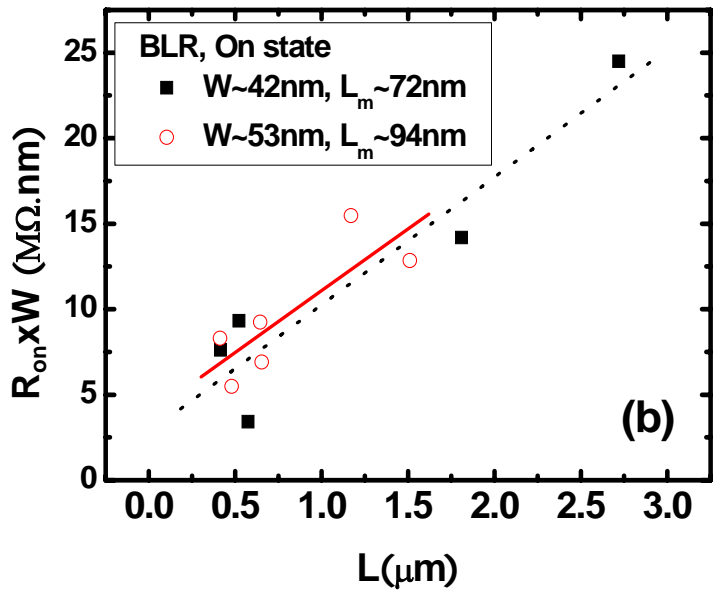

Xu Figure 3

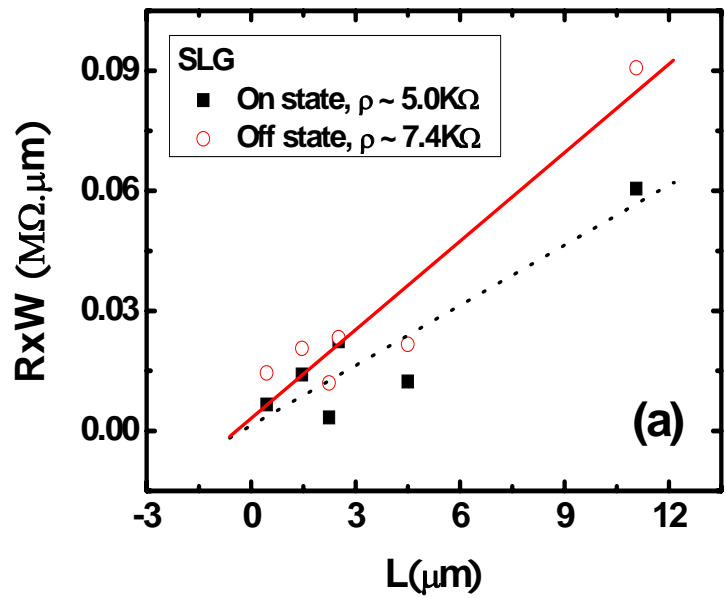

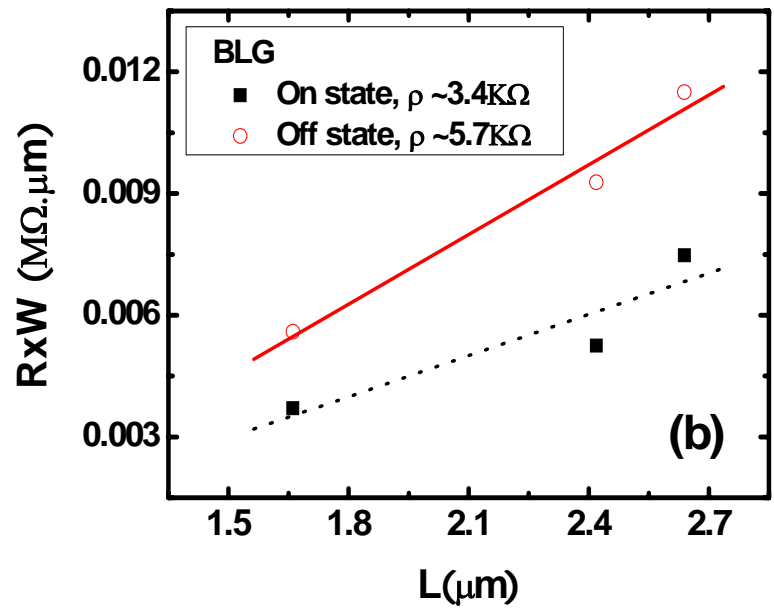

Xu Figure 4

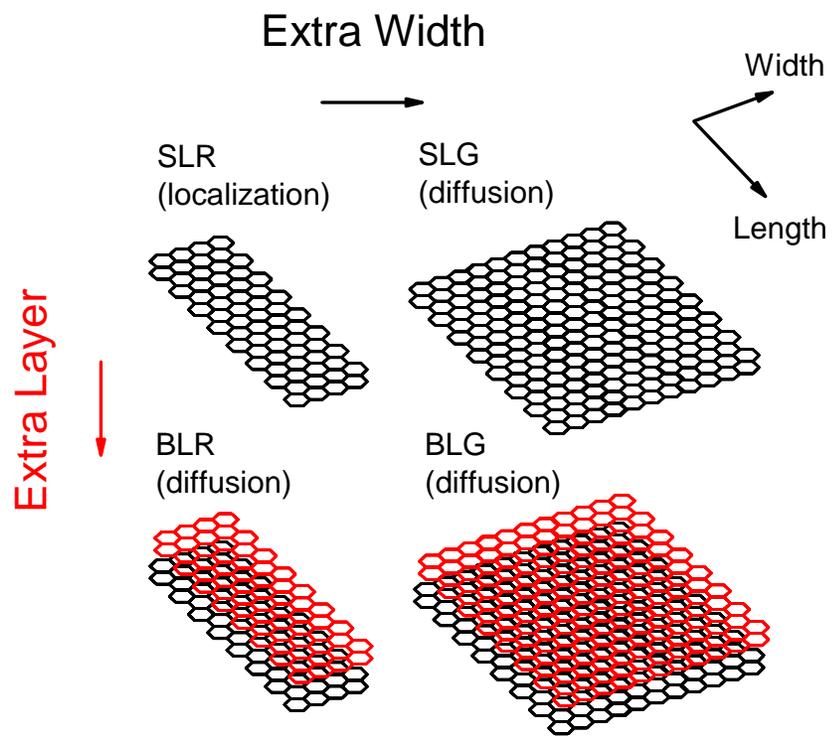

**Xu Figure 5**

# Edge effect on resistance scaling rules in graphene nanostructures

*Guangyu Xu, Carlos M. Torres Jr., Jianshi Tang, Jingwei Bai, Emil B. Song, Yu Huang, Xiangfeng Duan, Yuegang Zhang and Kang L. Wang*

**Supporting Materials**

<u>**Graphene sheet sample preparation (large width)**</u>

In this work, a thermally grown $SiO_2$ layer on a highly doped Si wafer is used as the bottom gate dielectric for the graphene devices. The thickness of the $SiO_2$ layer varies from 301 nm to 326 nm for different device batches. After the surface of $SiO_2$ is cleaned by acetone, isopropanol and oxygen plasma, graphene sheets are exfoliated and placed onto the $SiO_2$ substrate from the natural graphite flakes. The graphene sheets are characterized by optical microscopy and Raman spectroscopy. Then, a MMA/PMMA based dual-layer spin coating is applied followed by a 2 minute $150^\circ C$ baking. After an e-beam lithography step, a Titanium/Gold metal layer is evaporated to serve as the electrical contacts with a 7nm/80nm thickness, respectively.

The yield of this method exceeds 80% out of more than 40 fabricated graphene sheet devices (See Ref. S1).

<u>**Graphene nanoribbon sample preparation**</u>

Similarly, bulk graphene sheets are mechanically exfoliated from natural graphite and transferred onto a 300 nm thermally grown $SiO_2$ dielectric film on highly doped Si substrates, and are identified through optical microscopy and Raman spectroscopy. Then, silicon nanowires are deposited onto the graphene sheets to serve as an etch mask (diameter ranging from 20nm-50nm) and located by means of an optical microscope. A 30s oxygen plasma process is applied to etch away the exposed graphene, preserving the part of graphene (GNR) beneath the nanowires. The nanowires are removed by ultra-sonication, thus leaving only the GNR on the substrate. Finally, the GNRs are patterned to multi-probe structures. An MMA/PMMA based dual-layer spin coating is applied followed by a 2 minute $150^{\circ}C$ baking. After an e-beam lithography step, a Titanium/Gold metal layer is evaporated to serve as the electrical contacts with a thickness of 7nm/80nm, respectively.

The yield of this method exceeds 90% out of more than 70 fabricated GNR devices (see Ref. S2).

## Resistance measurement

The electrical measurement is performed in a Janis 500 four-arm probe station, pumped down to a vacuum of $10^{-6}$ torr at room temperature. A copper sample stage is used as the back-gate and contacted by the cold finger. The graphene and GNR devices are fixed onto the stage through the double-sided copper tapes. The gate bias is added through the copper tape to the highly doped Si, where the series resistance is generally below $1k\Omega$. An Agilent 4156C is used to apply dc bias to the device and to measure its dc resistance using MEDIUM integration time. During the measurement, the vacuum change is less than 3%.

All graphene and GNR devices are maintained in vacuum environment to avoid contact oxidation and uncontrollable doping effects from the ambiance. A 20 minute $100^{\circ}C$ vacuum bakeout process is generally applied to partially desorb contaminants. With the gate biases ranging from -100V to 100V, all devices show a gate leakage current generally below 4nA at room temperature.

By comparing the data from two- and four-terminal measurements, we find that the contact resistance ($R_c$) in SLR/BLR is small compared to the sample resistance R ($R_c \sim 10^3$-$10^4 \Omega \ll $ R), whereas $R_c \sim 10^3 \Omega < 0.5$ R in SLG/BLG.

## Sample dimension (length and width) measurement

Using VEECO Dimension 5000 atomic force microscopy (AFM) scanning, the width of GNRs is defined as the distance between the half-height points of the two edge profiles of GNRs (error<3nm from multiple measurements); the length of GNRs is measured from the AFM height image by scanning parallel to the metal electrodes to avoid the metal-edge effect (the shadow area would appear if the AFM scanning direction is not parallel to the metal). The error of length is typically less than 10nm from multiple measurements.

Using high magnification (50X) optical microscopy image with calibrated scale bars, we measured the length and width of large graphene sheet typically with an error less than 100nm from multiple measurements.

As such, each resistance data point in Fig. 2 through Fig. 4 is collected from an individual GNR device or an individual graphene sheet device.

**Length-dependence of $R_{off}/R_{on}$ in four graphene nanostructures**

The resistance scaling rules discussed in this work can also help us investigate the scaling behavior of other attributes in graphene nanostructures. As an example, Figure S1 shows the length-dependence of $R_{off}/R_{on}$ ratio for the four graphene nanostructures (SLR/BLR/SLG/BLG). No apparent features can be observed from these length-dependences of the $R_{off}/R_{on}$ ratio at room temperature. However, we roughly see that the $R_{off}/R_{on}$ ratios in SLR can be overall larger than that of BLR, whereas the $R_{off}/R_{on}$ ratios in SLG can be overall larger than that of BLG. This phenomenon is consistent with Y. Sui and J. Appenzeller's work[S3], which shows that the $I_{on}/I_{off}$ ratio decreases as the graphene thickness increases. The smaller $R_{off}/R_{on}$ ratio in our BLRs and BLGs may relate to the screening effect and interlayer coupling in bilayer graphene structures[S3].

Enhanced Conductance Fluctuation by Quantum Confinement Effect in Graphene Nanoribbons. *Nano Lett.* **10**, 4590-4594 (2010)

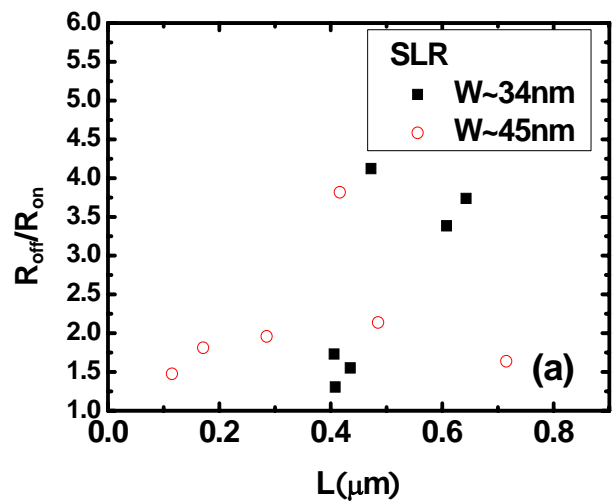
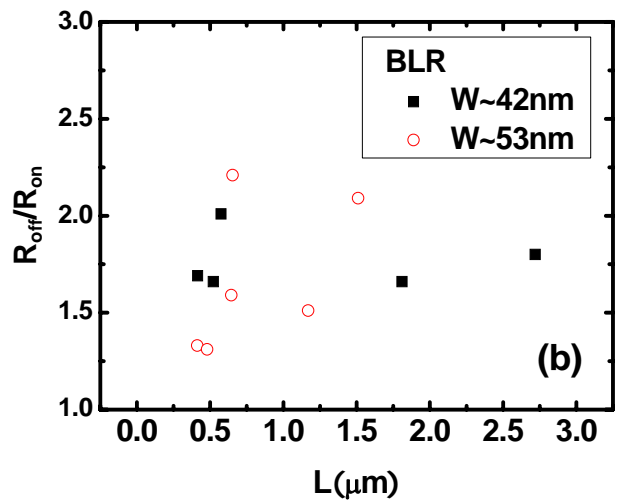
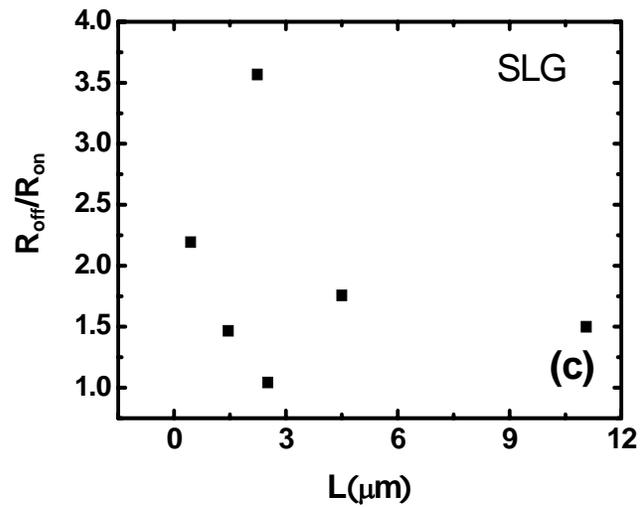
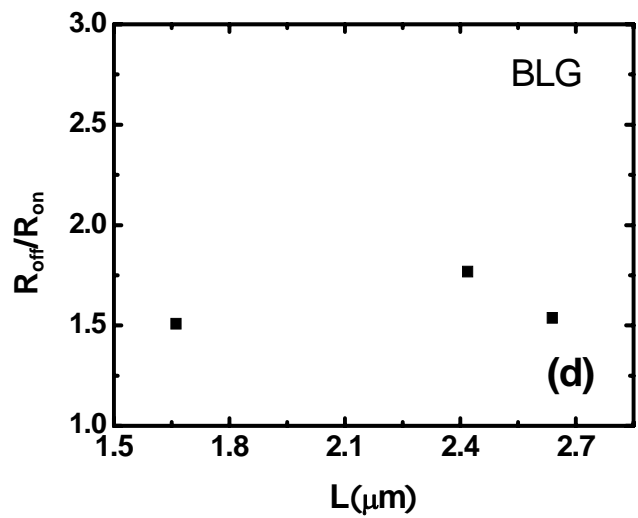

Xu Figure S1